\documentstyle[seceq, epsf, preprint]{ptptex}
\preprintnumber{KYUSHU-HET 46}

\title{Mass Hierarchy from \mathversion{bold}$SU(1,1)$ Horizontal Symmetry}
\author{ Kenzo INOUE\footnote{ e-mail address: inou1scp@mbox.nc.kyushu-u.ac.jp} and Nao-aki YAMASHITA\footnote{ e-mail address: naoaki@higgs.phys.kyushu-u.ac.jp}}

\inst{ Department of Physics, Kyushu University, Fukuoka 812-8581, Japan}
\abst{ 
The new mechanism for the generation of the chiral generations is proposed
based on the $SU(1,1)$ horizontal gauge symmetry.
The appearance of the chiral generations is controlled by the coupling constants of the model. 
This is crucial in the grand unification scheme.
The resulting chiral generations naturally acquire the hierarchal Yukawa coupling matrices.
Propriety of the model to the observed hierarchical structure is discussed numerically under the minimal parameter set.
}
\begin{document}

\maketitle
  
\section{Introduction}

  In the low-energy particle physics, Nature shows two remarkable aspects. 
First, quarks and leptons appear with the repetition of three generations.
It is not yet clear enough  where the definite number of chiral generations 
comes from.  
Second, the generations have the hierarchical mass structure.
This structure is translated to the hierarchical Yukawa coupling
structure whose origin is not yet well established.  
The characteristic structure of Yukawa coupling matrix seems to suggest some  inter-generation
symmetry, that is, horizontal symmetry.\cite{rf:horizon} 

For the possible approach to understanding these problems,
the model was proposed based on the noncompacet horizontal gauge symmetry $G_{H} = SU(1,1)$. \cite{rf:KI1} \cite{rf:KI2}
This model is a vector-like\cite{rf:RL} generalization\cite{rf:KAZUO} of the minimal supersymmetric standard model(MSSM)\cite{rf:MSSM}, 
and the three chiral generations and the hierarchical Yukawa coupling structure are realized  through the mechanism
called ``spontaneous generation of generations''.\cite{rf:KI1}

In the model, a MSSM chiral superfield($f$) is extended to a chiral superfield($F$) which belongs to an infinite dimensional unitary representation of the horizontal gauge symmetry $SU(1,1)$.
For example, the three generations of quark doublets $q$ is extended to a $SU(1,1)$ multiplet;
\begin{equation}
Q_{\alpha}= \{ q_{ \alpha }, q_{ \alpha + 1 }, q_{ \alpha + 2 }, q_{ \alpha + 3 }, \cdots \}, 
\end{equation}
where  $\alpha$ is the lowest weight of $SU(1,1)$ and assumed to be real positive. 
The three chiral $q$'s are contained in $Q_{\alpha}$. 
Precisely speaking, $q$'s are realized as linear combinations of infinite number of $q_{\alpha + i}$'s.
According to the vector-like(left-right) symmetry, all $F$'s are accompanied by $\bar{F}$'s which are conjugate to $F$'s under  $SU(1,1)\times SU(3) \times SU(2) \times U(1)$. 
So $Q_{\alpha}$ has a partner 
;
\begin{equation}
 \bar{Q}_{-\alpha} =   \{ \bar{q}_{ -\alpha }, \bar{q}_{ -\alpha - 1 }, \bar{q}_{- \alpha - 2 },
        \bar{q}_{ - \alpha - 3 }, \cdots \}. 
\end{equation}

  The horizontal symmetry is spontaneously broken by a non-vanishing vacuum expectation value(VEV)  of some component $\psi_{i}$ of $\Psi_{R}$, which is a multiplet belonging to a finite dimensional non-unitary representation of $SU(1,1)$ with the integer highest weight $R$,
\begin{equation}
\Psi_{R} = \{ \psi_{-R}, \psi_{-R+1}, \cdots , \psi_{R -1} , \psi_{R} \},
\end{equation}
 and assumed to be singlet under the standard model(SM) gauge group.

The coupling of the non-vanishing VEV $\left< \psi_{ -3} \right>$ to $F_{\alpha}$ and $\bar{F}_{- \alpha}$ provides the three chiral generations of $f$'s.
The relevant part of the superpotential taken in Reference 2) is 
\begin{eqnarray}
  M_{F}\bar{F}_{- \alpha} F_{\alpha} &+& v_{F} \Psi_{R} \bar{F}_{-\alpha} F_{\alpha} \nonumber \\ 
 &=& M_{F}\sum_{i=0}^{\infty}(-1)^{i}\bar{f}_{ -\alpha - i} f_{\alpha + i} + 
 v_{F}\sum_{i,j=0}^{\infty}A^{\alpha,R}_{i,j} \psi_{i-j}\bar{f}_{ - \alpha - i}f_{\alpha + j }, \label{eq:KI1}
\end{eqnarray}
where $M_{F}$ is a $SU(1,1)$ invariant mass and the Clebsch-Gordan coefficient 
$A_{i,j}^{\alpha,R}$ are given, with the normalization factor $N_{A}$, as\footnote{We correct the type miss of the equation (A8) in Reference 2).} 
\begin{eqnarray}
    A^{\alpha,R}_{i,j} &=& N_{A}(-1)^{j} \sqrt{ \frac{ i!j!(i-j + R)!(-i+j+R)!}
{\Gamma( 2 \alpha +i )\Gamma( 2 \alpha +j )}} \nonumber \\
 &  & \times
      \sum_{r= 0}^{ i-j +R}\frac{\Gamma( 2 \alpha +j +r)}{
  r!(j-i + r)!(R - r)!( i-j + R -r)!(r+j-R)!}. \label{eq:CG1}
\end{eqnarray}

When the $SU(1,1)$ is spontaneously broken and $\Psi_{R}$ is replaced with its VEV, the second term of (\ref{eq:KI1}) does not contain the first three components of $F_{\alpha}$.  
Under the presence of  the first term of (\ref{eq:KI1}), 
this absence of three components is retained in the unitary transformed way due to the fact that the second term of (\ref{eq:KI1}) is always dominant for large $i$ and $j$. 
This disappearance means a generation of the three chiral generations. 
The other components acquire huge Dirac masses with all components of $\bar{F}_{-\alpha}$ and decouple from the low-energy phenomena.    
The SM Yukawa couplings are obtained by extracting these three chiral generations from the $SU(1,1)$ invariant Yukawa couplings. 
The mixing effect due to the unitary transformation mentioned above and group theoretical structure of $SU(1,1)$ Clebsch-Gordan coefficients reproduce the hierarchical structure of Yukawa coupling matrices.

At the high-energy scale, where the horizontal symmetry is manifest, 
it is natural to expect that the SM gauge group is also unified into the higher symmetry(grand unified) group $G\supset SU(3)\times SU(2)\times U(1)$. \cite{rf:GUT}
Thus, it would be reasonable to extend the present model to the model based on the gauge group $SU(1,1) \times G$.
Grand unified theory(GUT) usually requires us to introduce extra particles which acquire super heavy masses through spontaneous break down of $G$.
The present model is basically vector-like, so that it has a desirable potential for making such extra particles super heavy. 
However the ``spontaneous generation of generations'' mechanism based on the superpotential (\ref{eq:KI1}) encounters serious difficulty under the grand unification. 
This is because the extra particles, such as colored Higgses, are unified with the MSSM particles into the same grand unified multiplet,
and consequently the chiral realization of the MSSM particles is always accompanied with that of the GUT partners.
Therefore we must abandon  (\ref{eq:KI1}) as the ``spontaneous generation of generations'' mechanism and search for an available alternative.

In this paper we propose the solution to this problem by a simple extension. 
The extension is to replace the $SU(1,1)$ invariant mass term $M_{F} \bar{F}_{- \alpha} F_{\alpha}$ in (\ref{eq:KI1}) by a Yukawa coupling $u_{F} \Phi_{R} \bar{F}_{- \alpha} F_{\alpha}$ with a non-unitary multiplet $\Phi_{R}$. 
Under this replacement, the vacuum has various phase structure.
When $v_{F} \left< \Psi \right>$ is much larger than $u_{F} \left< \Phi \right>$, 
the chiral structure is determined by  $\left< \Psi \right> \equiv \left<  \psi_{-g} \right>$ and we have $g$ chiral generations. 
On the contrary, when $v_{F} \left< \Psi \right> \ll u_{F} \left<  \Phi \right>$, we have no chiral generations of $F$, if $\left<  \Phi \right> \equiv \left<  \phi_{0} \right>$.
The number of chiral generations is integer which cannot vary continuously.
This means that the phase space of the vacuum is divided into two domains whose
boundary is characterized by the critical values of the ratio $u_{F} \left<  \Phi \right>/ v_{F} \left< \Psi \right>$.
This novel mechanism enables us to give heavy masses to superfluous particles like colored Higgses in the supersymmetric $SU(5)$ model\cite{rf:SUSYGUT} without fine tuning.
The purpose of this paper is to present a general framework of the model and to clarify the characteristics which the model reveals.

\section{The New Mechanism for Generating Chiral Generations}\label{genecon}

 Introducing the horizontal gauge symmetry $SU(1,1)$ in a vector-like manner, we extend the MSSM superfields $( q, \bar{u}, \bar{d}, l, \bar{e}, h, h' )$ and their conjugates $( \bar{q}, u, d, \bar{l}, e, \bar{h}, \bar{h'})$ to\cite{rf:KI1} 
\[ 
\begin{array}{ccccccc}
Q_{\alpha},& \bar{U}_{\beta},& \bar{D}_{\gamma},& L_{\eta},&\bar{E}_{\lambda},&
H_{ - \rho},& H'_{-\sigma},\\
\bar{Q}_{ - \alpha},& U_{ -\beta},& D_{ - \gamma},& \bar{L}_{ -\eta},&E_{ -\lambda},&
\bar{H}_{  \rho},& \bar{H'}_{\sigma},
\end{array}
\]
where each of the fields belong to an infinite dimensional representation of $SU(1,1)$ and $\alpha, \beta, \gamma, \eta, \lambda, \rho, \sigma$ are real positive. 

 Let us discuss the chiral structure of quark doublets $Q_{\alpha}$ and $\bar{Q}_{-\alpha}$.  We examine the superpotential 
\begin{equation}
 u_{Q}\Phi_{S}\bar{Q}_{-\alpha}Q_{\alpha} + 
   v_{Q}\Psi_{R}\bar{Q}_{- \alpha}Q_{\alpha},  \label{eq: KN1} 
\end{equation} 
where $v_{Q}$ and $u_{Q}$ are real coupling constants and  $\Psi_{R}$ and $\Phi_{S}$ belong to  finite dimensional representations of $SU(1,1)$ with highest weight $R$ and $S$ respectively.
We assume both $\Psi_{R}$ and $\Phi_{S}$ acquire non-vanishing VEV's $\left<\psi_{-g} \right>$ and $\left< \phi_{g'} \right>$ simultaneously. 
We must set $g'=0$ to give both possibilities that all modes become massive and $g$ modes become massless. 
Thus, the $SU(1,1)$ broken superpotential(mass terms) takes a form as
\begin{eqnarray}
u_{Q} \left< \phi_{ 0 } \right> 
     \sum_{i  = 0}^{\infty}A^{\alpha,S}_{i ,i} 
   \bar{q}_{ -\alpha - i} q_{  \alpha  + i }
+v_{Q} \left< \psi_{ -g } \right> \sum_{i = 0}^{\infty}
      A^{\alpha,R}_{i,i + g}\bar{q}_{ - \alpha - i} q_{ \alpha + i + g }.   \label{eq:Broken-NK}   
\end{eqnarray} 
The generation of chiral generations means the appearance of massless modes in
(\ref{eq:Broken-NK}).  

Let us assume that we have $g$ massless modes $\{ q^{(0)}, q^{(1)}, q^{(2)},\cdots,q^{(g-1)} \}$ and write
\begin{equation}
 q_{\alpha + j} = \sum_{ i=0 }^{g-1 }   a^{q(i)}_{j} q^{(i)} + \mbox{ massive modes}. 
\label{eq:unitary-tr}
\end{equation}
The coefficients $a^{q(i)}_{j}$ must be normalizable as $\sum_{j = 0}^{\infty} | a^{q(i)}_{j} |^{2} =1$.
Inserting (\ref{eq:unitary-tr}) into (\ref{eq:Broken-NK}), we obtain the massless condition for $q^{(i)}$ as 
\begin{equation}
  u_{Q} \left< \phi_{ 0 } \right>  A^{\alpha,S}_{j ,j} a^{q(i)}_{ j} + 
  v_{Q} \left< \psi_{ -g } \right> A^{\alpha,R}_{j,j + g} a^{q(i)}_{j + g}  = 0.\end{equation}
This equation gives recursion relation for coefficients $a^{q(i)}_{j}$, 
\begin{equation}
  a^{q(i)}_{j + g} 
   = - \frac{ u_{Q} \left< \phi_{ 0 } \right>  A^{\alpha,S}_{j ,j} }
      { v_{Q} \left< \psi_{ -g } \right> A^{\alpha,R}_{j,j + g} } a^{q(i)}_{j}. \label{eq:recursion-relation}
\end{equation}
This recursion relation means that $a^{q(i)}_{j}$ goes to zero if the magnitude of the coefficient of $a^{q(i)}_{j}$ in the right-hand side is smaller than unity in the limit  $j \rightarrow \infty$.
In this case $a^{q(i)}_{j}$ are normalizable, so massless modes appear.
  
For fixed $i-j$, the asymptotic behavior of the Clebsch-Gordan coefficient (\ref{eq:CG1}) is 
\begin{equation}
A^{\alpha, R}_{i,j} \sim  N_{A}(-1)^{j}\frac{(2R)!}{(R!)^{2}\sqrt{(R-i+j)!(R+j-i)!}}(j)^{R},~~~( i,~j \rightarrow \infty). 
\end{equation}
Therefore, we find, in the limit $j \rightarrow \infty$,  
\begin{equation}
  \frac{ u_{Q} \left< \phi_{ 0 } \right>  A^{\alpha,S}_{j ,j} }
      { v_{Q} \left< \psi_{ -g } \right> A^{\alpha,R}_{j,j + g} } \sim
-(-1)^{g} 
  \frac{ u_{Q} \left< \phi_{ 0 } \right>}{ v_{Q} \left< \psi_{ -g } \right>} 
   \frac{(R!)^{2}(2S)!}{(S!)^{2}(2R)!}\sqrt{ \frac{(R-g)!(R+g)!}{R!R!}}\frac{j^{S}}{j^{R}}. \label{eq:condition}
\end{equation}
For $S < R$, the right-hand side goes to zero. 
This means that all $a^{q(i)}_{j}$ are well defined,
so $g$ massless generations appear. 
By setting $g=3$, the three chiral generations of quark doublets are realized.
On the other hand, for $S > R $, $a^{q(i)}_{j}$ diverges for $j \rightarrow \infty$ and we can not define normalizable massless modes. This means that all components in (\ref{eq:Broken-NK}) become massive.
Obviously, the present model is a natural extension of the previous model based on (\ref{eq:KI1}), which is recovered by the choice $S=0$.
This extension, nevertheless, gives the model with the remarkable feature,
which is realized in the special case $S=R$.

 For $S=R$, the situation in (\ref{eq:Broken-NK}) is subtle. 
Since $j^{S}/j^{R}$ is unity, 
the coefficient of $j^{S}/j^{R}$ in the right-hand side of (\ref{eq:condition}) determines whether the chiral generations appear or not. 
Let us define the ratio $r_{Q}$ by
\begin{equation}
r_{Q} = 
\frac{ u_{Q} \left< \phi_{ 0 } \right>}{ v_{Q} \left< \psi_{ -g } \right>} . \label{eq:ratio}
\end{equation}
The asymptotic behavior of $a^{q(i)}_{j}$ is governed by the value of this ratio.
From (\ref{eq:recursion-relation}) and  (\ref{eq:condition}) we find the critical value of $r_{Q}$ is given by
\begin{equation}
r^{c}_{g} = \sqrt{ \frac{R!R!}{(R-g)!(R + g)!}}. \label{eq:g-cri} 
\end{equation}
For $|r_{Q}| <  r^{c}_{g}$, $a_{j}^{q(i)}$ is normalizable and the $g$ chiral generations appear. The chiral generation structure does not appear for $|r_{Q}| \geq  r^{c}_{g}$. 
The generation of generations depends on the magnitude of $r_{Q}$ which is controlled by couplings $u$ and $v$. 
That is, two dimensional parameter space of $u$ and $v$ is divided into two domains, each of which has the quite different chiral structure. 

The choice $S=R$ not only accomplishes simplification purpose, but also gives us an attractive mechanism.
  Some of the MSSM particles are unified in a large GUT multiplet with extra particles at high energy.
If each particles in the GUT multiplet have different ratio $r$'s (\ref{eq:ratio}), it is possible to realize without fine tuning that the MSSM particles have the chiral structure and the extra particles are super heavy.

For instance, let us consider the Higgs sector of the supersymmetric $SU(5)$ grand unified model.\cite{rf:SUSYGUT}
A Higgs doublet $h$ is embedded in a $\mathbf{5}$-dimensional representation together with an extra colored Higgs $h^{C}$. 
The $SU(1,1)$ Higgs multiplet is denoted as
\begin{equation}
K_{ - \rho} = \left( \begin{array}{c} H^{C}_{ -\rho} 
                 \\ H_{ - \rho} \end{array} \right).
\end{equation}
We introduce two multiplet  $\Phi'_{R'}$ and ${\Psi'}_{R'}$ which belong to  $\mathbf{1}$ and $\mathbf{24}$ of $SU(5)$ respectively, and couple them to $K_{ - \rho }$ and its conjugate $\bar{K}_{  \rho }$ by
\begin{equation}
 u_{K}\Phi'_{R'} \bar{K}_{ \rho }  K_{ -\rho } 
  + v_{K}\bar{K}_{ \rho } \Psi'_{R'} K_{ - \rho }.
\end{equation}
The gauge symmetry $SU(1,1)\times SU(5)$ is spontaneously broken to $SU(3)\times SU(2)\times U(1)$ via non-vanishing VEV's of $\Phi'_{R'}$ and ${\Psi'}_{R'}$ 
\begin{equation}
 \left< \phi'_{0} \right> \neq 0, \qquad \left< \psi'_{1} \right> \left( \begin{array}{ccccc}
                            1 & & & & \\  & 1 & & & \\ & & 1 & & \\
                              & & &-\frac{3}{2} & \\  & & & & -\frac{3}{2} \\
                              \end{array} \right) \neq 0,
\end{equation}
where we have set $g=-1$ for $\Psi'_{R}$ to realize one chiral generation for Higgs because of its negative weights of $SU(1,1)$.
The ratios (\ref{eq:ratio}) for $h$ and $h^{C}$ are
\begin{equation}
  r_{H^{C}}  = \frac{u_{K} \left< \phi'_{0} \right>}{ v_{K} \left< \psi'_{1} \right>} ,\label{eq:chiggs1}
\end{equation}
\begin{equation}
  r_{H}  = - \frac{2}{3} \frac{u_{K} \left< \phi'_{0} \right>}{ v_{K} \left< \psi'_{1} \right>}. \label{eq:chiggs2}
\end{equation}
The critical value in the present case is given by (\ref{eq:g-cri}) with $g = -1$,~$r^{c}_{1}$.
When the ratio $r$'s satisfies the condition $|r_{H}| < r_{1}^{c} \leq |r_{H^{C}}|$, $H$ has one chiral generation but $H^{C}$ does not.
This condition is consistent with (\ref{eq:chiggs1}) and  (\ref{eq:chiggs2}).
Therefore, we can remove the colored Higgs in a low-energy theory by choosing such $r$'s. There is no fine tuning. 
Like this example, we can use this mechanism to make superfluous particles heavy.

Following above discussion, we describe the generation generating superpotential as 
\begin{eqnarray}
(u_{Q}\Phi_{R} &+&  v_{Q}\Psi_{R})\bar{Q}_{-\alpha}Q_{\alpha} +
(u_{U}\Phi_{R} +  v_{U}\Psi_{R})U_{-\beta} \bar{U}_{\beta} +
(u_{D}\Phi_{R} +  v_{D}\Psi_{R})D_{-\gamma} \bar{D}_{\gamma}  \nonumber \\
&+&
(u_{L}\Phi_{R} +  v_{L}\Psi_{R})\bar{L}_{-\eta}L_{\eta} +
(u_{E}\Phi_{R} +  v_{E}\Psi_{R})E_{-\lambda}  \bar{E}_{\lambda} \nonumber \\
&+&
(u_{H}\Phi'_{R'} +  v_{H}\Psi'_{R'}) \bar{H}_{\rho}H_{-\rho} +
(u_{H'}\Phi'_{R'} +  v_{H'}\Psi'_{R'}) \bar{H'}_{\sigma}H'_{-\sigma}.   \label{eq:GSP}
\end{eqnarray}
For simplicity and economical purpose, 
we assume the same non-unitary multiplets $\Phi_{R}$ and $\Psi_{R}$ couple to quarks and leptons and generate the three chiral generations through non-vanishing VEV's $\left<\phi_{0}\right>$ and $\left<\psi_{-3} \right>$.
Also, the same $\Phi'_{R'}$ and $\Psi'_{R'}$ with  $\left<\phi'_{0} \right> \neq 0$ and $\left<\psi'_{1} \right> \neq 0$ realize one chiral generation for Higgses, 
The potential (\ref{eq:GSP}) allows the selective
generation of the chiral generations and the extension to grand unified models.

\section{The Hierarchical Structure of Yukawa Coupling Matrices}

Let us now proceed to the Yukawa couplings of Higgses to quarks and leptons.
The relevant terms in the superpotential are given as
\begin{equation}
y_{U} \bar{U}_{\beta}Q_{\alpha}H_{-\rho} +  y_{D} \bar{D}_{\gamma}Q_{\alpha}H'_{-\sigma} + y_{E} \bar{E}_{\lambda}L_{\eta}H'_{-\sigma} . 
\label{eq:yukawa-sector}
\end{equation}
The $SU(1,1)$ invariance requires 
\begin{eqnarray}
\rho &=& \alpha + \beta + \Delta, \\
\sigma &=& \alpha + \gamma + \Delta' = \lambda + \eta + \Delta'',
\end{eqnarray}
where $\Delta$, $\Delta'$ and $\Delta''$ are non-negative integers.
The Clebsch-Gordan decomposition of the first term of (\ref{eq:yukawa-sector})
takes the form
\begin{equation}
y_{U}\sum_{i,j=0}^{\infty} B^{U}_{i,j}\bar{u}_{\beta + i }q_{\alpha + j}
h_{-\rho - i -j + \Delta}. \label{eq:up-yukawa}
\end{equation}
The Clebsch-Gordan coefficient is
\begin{eqnarray}
B^{U}_{i,j} =&& N^{U}(-1)^{i+j}
\sqrt{ \frac{ i!j! \Gamma(2\beta + i)\Gamma(2\alpha + j)}{
( i + j  - \Delta )!\Gamma(2\rho + i + j - \Delta)}} \nonumber \\
   &&\times \sum_{r = 0}^{ \Delta} 
     \frac{ (-1)^{r}( i+j-\Delta )!}{ (i -r)!(j - \Delta + r)! r!(\Delta -r)!
\Gamma(2\beta + r)\Gamma(2\alpha + \Delta - r)},
\end{eqnarray}
where $N^{U}$ is a normalization constant.

To find the structure of Yukawa couplings for massless modes, 
we must pick up them from original components. 
As mentioned in section \ref{genecon}, the massless modes of $Q_{\alpha}$ which consist of  $g=3$ chiral generations
are given from (\ref{eq:unitary-tr}) as
\begin{equation}
q^{(i)} = \sum_{j = 0}^{\infty}{a^{q(i)}_{j}}^{*}q_{\alpha + j} \qquad (i = 0, 1, 2). \label{eq:tr1}
\end{equation}
The coefficients $a^{q(i)}_{j}$ are determined by the equation (\ref{eq:recursion-relation}) as 
\begin{equation}
a^{q(i)}_{3k+i} = (r_{Q})^{k}a^{q(i)}_{i} \prod_{n=0}^{k-1}\frac{ A^{\alpha,R}_{3n+i ,3n+i} }{ A^{\alpha,R}_{3n+i ,3(n + 1)+i} }, \label{eq:cef1}
\end{equation}
where
\begin{equation}
  r_{Q} = \frac{ u_{Q} \left< \phi_{ 0 } \right> }{ v_{Q} \left< \psi_{ -3 } \right>}.
\end{equation}
To guarantee that the three chiral generations of quark doublets are realized, we assume that the ratio $r_{Q}$ satisfies the relation 
\begin{equation}
 |r_{Q}| < r^{c}_{3}= \sqrt{\frac{R(R -1)(R -2 )}{(R+1)(R+2)(R+3)}}. \label{eq:three-gene}
\end{equation}

The mixing contribution of higher components $q_{\alpha + j}$ to massless modes is generally suppressed, for reasonable value of $|r_{Q}| < 1$, due to the 
suppression factor $(r_{Q})^{k}$ in (\ref{eq:cef1}). 
There exists an additional suppressing mechanism which becomes effective for large value of $R$. 
From the asymptotic form of Clebsch-Gordan coefficient $A^{\alpha,R}_{i,j}$ in the limit $R\rightarrow \infty$,
\begin{equation}
A^{\alpha,R}_{i,j} \sim N_{A}(-1)^{j} \frac{1}{\sqrt{i!j!\Gamma( 2 \alpha + i)
\Gamma( 2 \alpha + j)}}R^{2\alpha + i + j -1},
\end{equation}
we find the suppression factor
\begin{equation}
a^{q(i)}_{3k+i} \propto (r_{Q})^{k} a_{i}^{q(i)} \left(\frac{1}{R^{3}}\right)^{k}.  
\end{equation}

 For all quarks and leptons which have the three chiral generations, 
we have similar relations (\ref{eq:tr1})-(\ref{eq:three-gene}) with appropriate replacements of the weights and couplings $u$, $v$.

Higgses have negative weights so that the Yukawa couplings (\ref{eq:yukawa-sector}) are invariant under $SU(1,1)$. 
We can obtain one massless mode by setting the non-vanishing VEV for positive component of $\Psi'$ as $\left<\psi'_{1}\right> \neq 0$.
So the massless mode of $H_{- \rho}$ is given as
\begin{equation}
h = \sum_{j = 0}^{\infty}{a^{h}_{j}}^{*}h_{-\rho - j},\label{eq:h-mix} 
\end{equation}
where
\begin{equation}
  a^{h}_{j} = (r_{H})^{j}a_{0}^{h}\prod_{n=0}^{j-1}\frac{ A^{-\rho,R'}_{n ,n} }{ A^{-\rho,R'}_{n+1 ,n} }, \label{eq:h-mix-form}
\end{equation}
\begin{equation}
r_{H} = \frac{ u_{H} \left< \phi'_{ 0 } \right> }{ v_{H} \left< \psi'_{ 1 } \right>}.
\end{equation}
Here we assume 
\begin{equation}
 |r_{H}| < r^{c}_{1} = \sqrt{ \frac{R'}{R'+1}},
\end{equation}
to realize one chiral generation. Similar relations apply to $H'_{-\sigma}$.
The mixing effect in (\ref{eq:h-mix}) is also suppressed for $|r_{H}| < 1$ and we have additional suppression in (\ref{eq:h-mix-form}) for large value of $R'$,
\begin{equation}
a^{h}_{j} \propto (r_{H})^{j}a^{h} \left( \frac{1}{R'} \right)^{j}. 
\end{equation}

From the $SU(1,1)$ invariant couplings (\ref{eq:yukawa-sector}), 
we can extract Yukawa couplings among massless modes.
For example, for up-type quarks, we have
\begin{equation}
\sum_{i,j=0}^{2} \Gamma^{u}_{ij}  \bar{u}^{(i)}  q^{(j)} \ h, 
\end{equation}
where the Yukawa coupling matrix $\Gamma^{u}$ is given as 
\begin{equation}
\Gamma^{u}_{ij} = y_{U} \sum_{m,n = 0}^{\infty} 
B^{U}_{3m+i, 3n+j}a^{u(i)}_{3m+i} a^{q(j)}_{3n + j}a^{h}_{3(m + n) + i + j - \Delta} 
. \label{eq:realYukawa}
\end{equation}
Other Yukawa coupling matrices $\Gamma^{d}$ and $\Gamma^{e}$ are obtained similarly.

This general formula (\ref{eq:realYukawa}) is very complex, and it is difficult to explore the general property analytically.
However, the suppressing effect of mixing in (\ref{eq:tr1}) and (\ref{eq:h-mix}) realizes the hierarchy among the generations generally.    
We can expand $\Gamma^{u}_{ij}$ in terms of the power series of $r$'s and extract the lowest order contribution of $r$'s. 
 We find that the phenomenologically attractive structure
is realized when we choose $\Delta = 0$.  In this case the leading
contribution to $\Gamma^{u}$ comes from the term with $m=n=0$ in
(\ref{eq:realYukawa}).  Omitting the numerical coefficients, it takes
the form
\begin{equation}
\Gamma^{u}(\Delta = 0) \propto
   \left( \begin{array}{ccc}
       1  & 
      r_{H} &
      r_{H}^{2} \\
      r_{H} &
      r_{H}^{2}&
      r_{H}^{3} \\
      r_{H}^{2} &
      r_{H}^{3} &
      r_{H}^{4}
      \end{array} \right).  \label{eq:ymatrix}
\end{equation}
The terms with $m > 0$ and/or $n > 0$ in (\ref{eq:realYukawa}) give the modification to the lowest order term with $m=n=0$. 
In this simplification, $r_{H}$ governs the matrix structure. 
The masses of each generation are $m_{0} \propto {\cal O}\left( 1 \right)$, $m_{1} \propto {\cal O}\left(  r_{H}^{2} \right)$ and  $m_{2} \propto {\cal O}\left( r_{H}^{4} \right)$. 

\section{Numerical Analysis Under the Minimal Parameter Set} \label{sec4}

In this section, we show the result of numerical analysis for the typical parameter region to see whether the observed hierarchical structure is realized or not.
There are many free parameters in the model; seven weights of unitary fields ($\alpha$, $\beta$, $\gamma$, $\eta$, $\lambda$, $\rho$, $\sigma$), two weighs of non-unitary field ($R$, $R'$), seven ratios of VEV's and couplings ($r_{Q}$, $r_{U}$, $r_{D}$, $r_{L}$, $r_{E}$, $r_{H}$, $r_{H'}$), and three Yukawa coupling constants ($y_{U}$, $y_{D}$, $y_{E}$).
In practice, it is impossible to make numerical analysis retaining all of this freedom. 
Based on the consideration of grand unification, we decrease the number of these parameters.

First of all, we assign the same lowest weight $\alpha$ to all quarks and leptons, $\alpha = \beta = \gamma = \eta = \lambda$. 
Then the weights of Higgses are $\rho= \sigma = 2\alpha$.
We further assume $R=R'$ and make the identification  $\Phi_{R} = \Phi'_{R'}$ in the potential (\ref{eq:GSP}).
To check whether our framework has capacity for describing the observed hierarchical structure,
we calculate mass ratios of quarks and leptons among the generations and the CKM matrix\cite{rf:CKM} under this parameter set. 
In the calculation of these quantities, we do not need values of Yukawa couplings $y_{U}, y_{D}, y_{E}$ and normalization constants of Clebsch-Gordan coefficients. 

\begin{figure}[bt]
\begin{center}
\setlength{\unitlength}{0.65cm}
\begin{picture}(10,7)
\epsfxsize=75mm
\put(0,0){\epsfbox{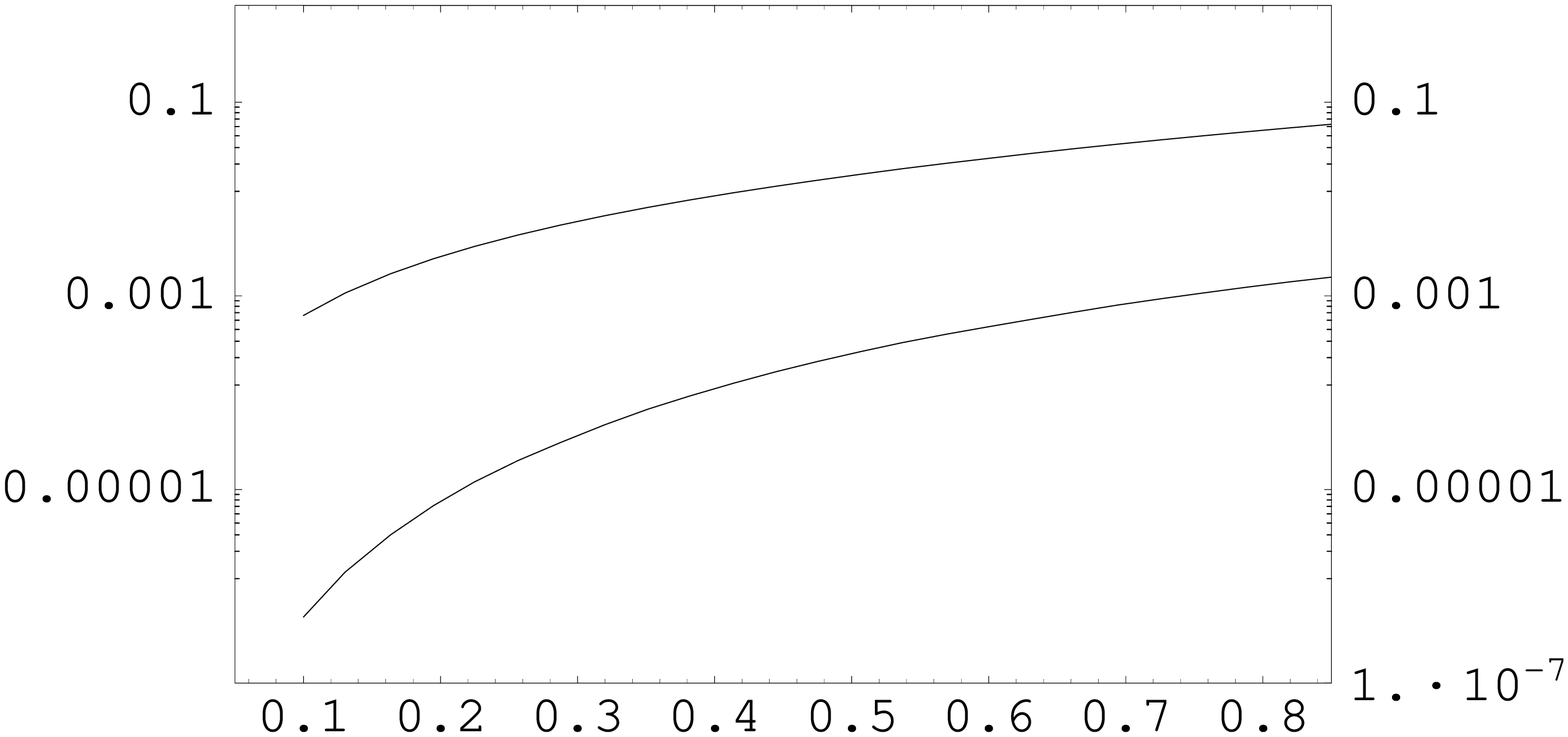}}
\put(4.5,0.3){\footnotesize $|r_{H}| ~(|r_{H'}|)$}
\put(5,5.3){\footnotesize $\frac{m_{1}}{m_{0}}$}
\put(6,3.2){\footnotesize $\frac{m_{2}}{m_{0}}$}
\end{picture}
\end{center}
\caption{The mass ratios in the case of $R= 3$, $\alpha  = 0.5$, $r_{Q} = 0$ and  $r_{U}\mbox{($r_{D}$)} = 0$}
\label{fig1}
\end{figure}

\begin{figure}[bt]
\begin{center}
\setlength{\unitlength}{0.65cm}
\begin{picture}(10,7)
\epsfxsize=75mm
\put(0,0){\epsfbox{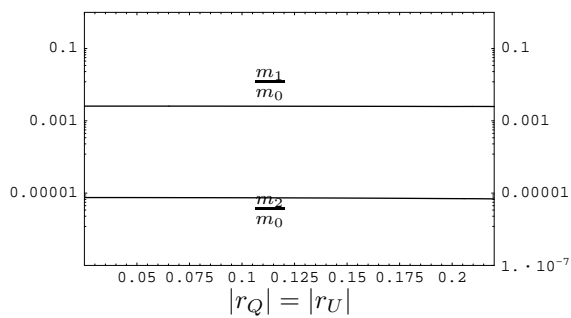}}
\put(4.5,0.3){\footnotesize $|r_{Q}| = |r_{U}|$}
\put(5,4.8){\footnotesize $\frac{m_{1}}{m_{0}}$}
\put(5,2.2){\footnotesize $\frac{m_{2}}{m_{0}}$}
\end{picture}
\end{center}
\caption{The mass ratios in the case of $R = 3$, $\alpha = 0.5$ and  $|r_{H}| = 0.2$}
\label{fig2}
\end{figure}

First, we set $\alpha = 0.5$ and $R=3$ which is minimal allowed value of $R$,  and  
investigate the dependence of the masses on the ratios $r$'s. 
In the case of $R=3$, to realize one generation for Higgses and three generations for quarks, 
we must restrict values of $r's$ to $|r_{H}|, |r_{H'}| < \sqrt{3}/ 2$ 
and $|r_{Q}|, |r_{U}|, |r_{D}| < 1 / \sqrt{20}$.

Fig. 1 shows  mass rations $ m_{1}/m_{0}$ and $m_{2}/m_{0}$ of up-type quarks(down-type quarks) for various values of $r_{H}$($r_{H'}$) 
when $r_{Q} =0$ and $r_{U}=0$($r_{D}=0$).
It is surprising that the resulting mass hierarchy is much larger than what is naively expected from (\ref{eq:ymatrix})
( $m_{1}/m_{0} \sim {\cal O}(r_{H}^{2})$, $m_{2}/m_{0} \sim {\cal O}(r_{H}^{4})$). 
In fact, the order of the observed mass ratios of up-type quarks, 
$m_{c}/m_{t} =  m_{1}/m_{0} \sim {\cal O}(10^{-3})$,
$m_{u}/m_{t} = m_{2}/m_{0} \sim {\cal O}(10^{-6})$, 
and down-type quarks, $m_{s}/m_{b} = m_{1}/m_{0} \sim {\cal O}(10^{-2})$, 
$m_{b}/m_{d} = m_{2}/m_{0} \sim {\cal O}(10^{-4})$),
 is realized by setting $r_{H} \approx 0.2 \sim 0.25$ and 
$r_{H'} \approx 0.4 \sim 0.65$.

In Fig. 2, we show the effect of non-vanishing values of $r_{Q}$ and $r_{U}$ with keeping $r_{Q}=r_{U}$.
Evidently the dependence of masses on the values of $r_{Q}$, $r_{U}$ and $r_{D}$ is small.
This situation has been vaguely anticipated in the end of the previous section. 

Generally the value of ratio $r$ is complex. 
From (\ref{eq:GSP})  we observe that $r_{Q}$, $r_{U}$ and $r_{D}$ have the common phase 
$\arg [ \left< \phi_{0} \right> / \left< \psi_{-3} \right>  ]$, and also $r_{H}$ and $r_{H'}$ have the phase $\arg [ \left< \phi_{0} \right> / \left< \psi'_{1} \right>  ]$.
The physically meaningful phase is $SU(1,1)$ invariant phase
\begin{equation}      
\theta = \arg \left[ \left<\psi_{-3} \right> \right] + 3 \arg \left[  \left<\psi'_{1} \right> \right]. \label{eq:phase-r}
\end{equation}      
The result of Fig. 2 (insensitivity to  $\left<\psi_{-3} \right>$) implies that the effect of this phase $\theta$ on masses is small.

\begin{figure}[bt]
\begin{center}
\setlength{\unitlength}{0.65cm}
\begin{picture}(10,7)
\epsfxsize=75mm
\put(0,0){\epsfbox{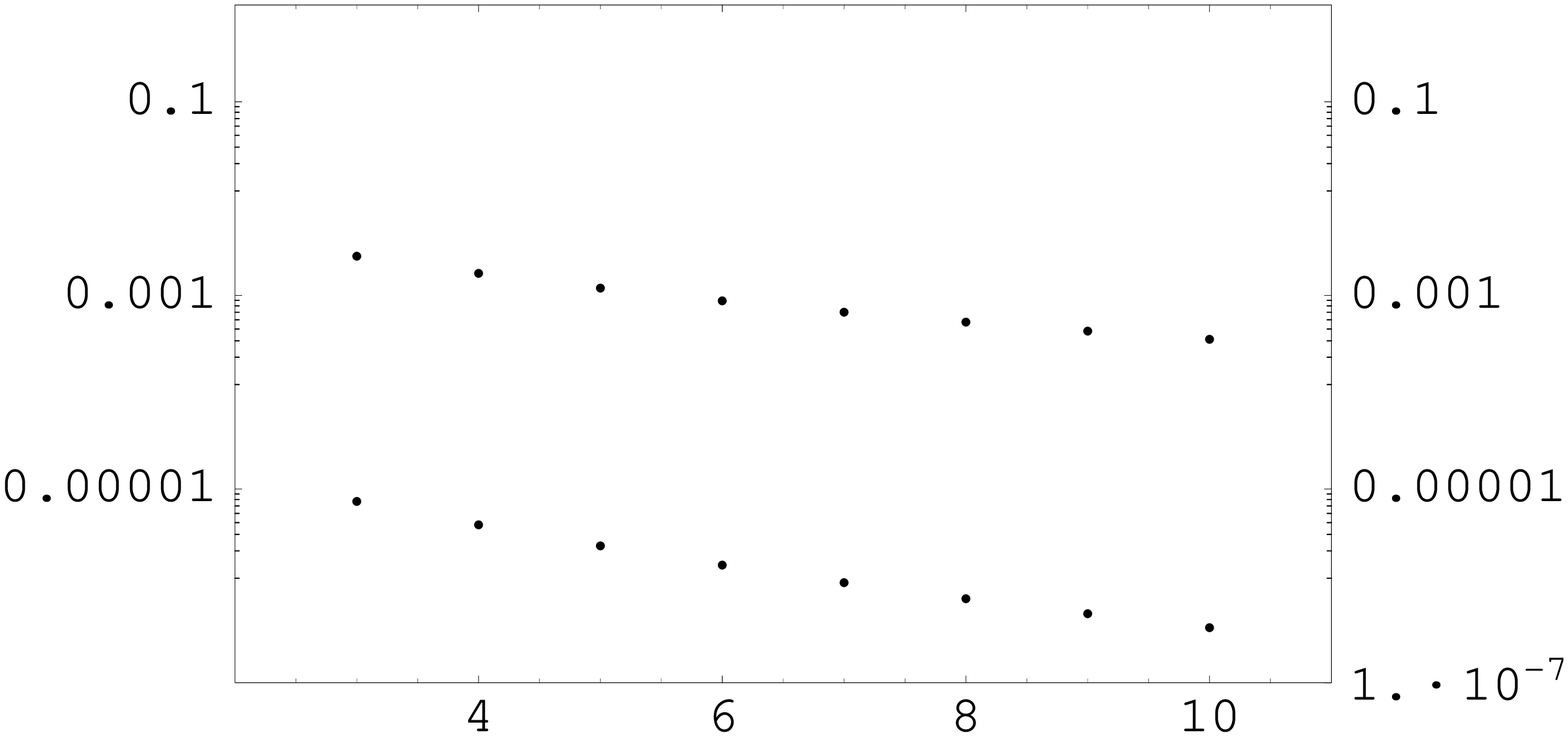}}
\put(5,0.3){\footnotesize $R$}
\put(4,4.7){\footnotesize $\frac{m_{1}}{m_{0}}$}
\put(6,2.5){\footnotesize $\frac{m_{2}}{m_{0}}$}
\end{picture}
\end{center}
\caption{The mass ratios  in the case of $\alpha  = 0.5$, $r_{H}= 0.2$, $r_{Q} = 0.1$ and $r_{U} = 0.1$}
\label{fig3}
\end{figure}

\begin{figure}[bt]
\begin{center}
\setlength{\unitlength}{0.65cm}
\begin{picture}(10,7)
\epsfxsize=75mm
\put(0,0){\epsfbox{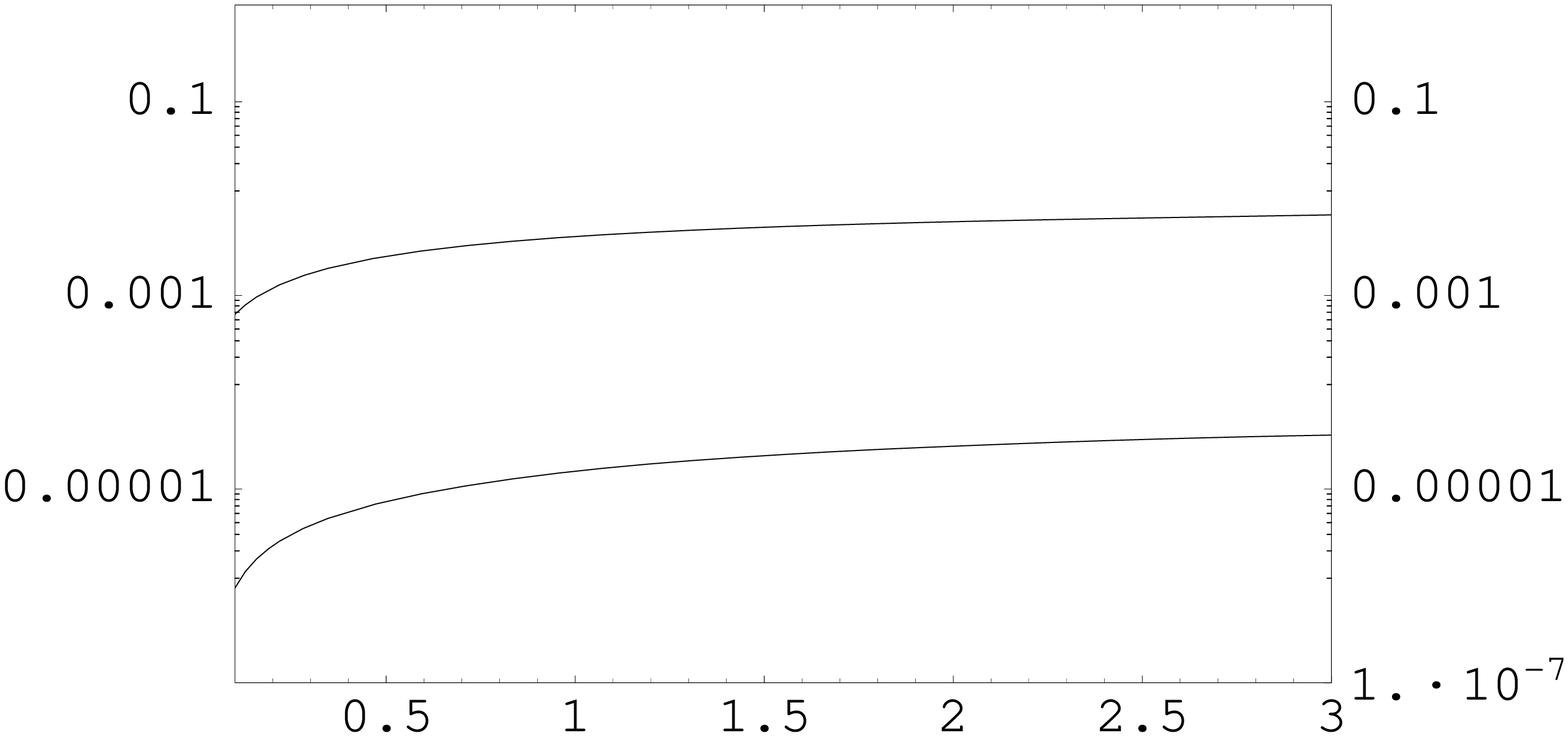}}
\put(5,0.3){\footnotesize $\alpha$}
\put(4,5){\scriptsize $\frac{m_{1}}{m_{0}}$}
\put(6,3.4){\scriptsize $\frac{m_{2}}{m_{0}}$}
\end{picture}
\end{center}
\caption{The mass ratios in the case of $R = 3$, $r_{H} = 0.2$, $r_{Q} = 0.1$ and $r_{U} = 0.1$}
\label{fig4}
\end{figure}

The $R$ dependence of mass ratios is given in Fig. 3. 
We see the mass hierarchy is enhanced slowly with an increase in the value of $R$.
The $\alpha$ dependence is given in Fig. 4. 
The variation of $\alpha$ gives sizable effect, although not so large.
We reproduce the observed mass hierarchy for quarks by choosing adequate values of $r_{H}$ and $r_{H'}$ under the minimal parameter set.

The hierarchical structure of Yukawa coupling matrices $\Gamma^{u}$ and $\Gamma^{d}$ reflects itself in the characteristic form of the CKM matrix.
Naively, the CKM matrix is expected to take the form
\begin{equation}
V_{\mbox{\tiny CKM}}
 \sim  
  \left( \begin{array}{ccc}
       1   & r  & r^{2}  \\
       r   & 1  & r  \\
       r^{2}  & r & 1  \end{array} \right).
\end{equation}
As a typical example, we give the numerical result in the case $R = 3$, $\alpha=0.5$, $r_{H} = 0.2$, $r_{H'} = 0.45$, $r_{Q} = 0.2$, $r_{U} = 0.2$ and $r_{D} = 0.2$; 
\begin{equation}
\left( \begin{array}{ccc}
     |V_{ud}| & |V_{us}| & |V_{ub}| \\
     |V_{cd}| & |V_{cs}| & |V_{cb}| \\
     |V_{td}| & |V_{ts}| & |V_{tb}|   \end{array} \right) 
 \sim  
  \left( \begin{array}{ccc}
       0.975   & 0.221  & 0.006  \\
       0.221   & 0.973  & 0.074  \\
       0.010  & 0.088 & 0.997 \end{array} \right).  \label{eq:CKMget1}
\end{equation}
The mass ratios in this case are $m_{c}/m_{t} = 2.5\times 10^{-3}$, $m_{u}/m_{t} = 7.0 \times 10^{-6}$, $m_{s}/m_{b} = 1.4\times 10^{-2}$ and  $m_{d}/m_{b} = 1.7\times 10^{-4}$.
Although these mass ratios are not enough satisfactory, it is rather surprising that such a simple choice of parameters reproduces the qualitative feature of $V_{\mbox{\tiny CKM}}$.

\begin{figure}[bt]
\begin{center}
\setlength{\unitlength}{0.65cm}
\begin{picture}(10,7)
\epsfxsize=75mm
\put(0,0){\epsfbox{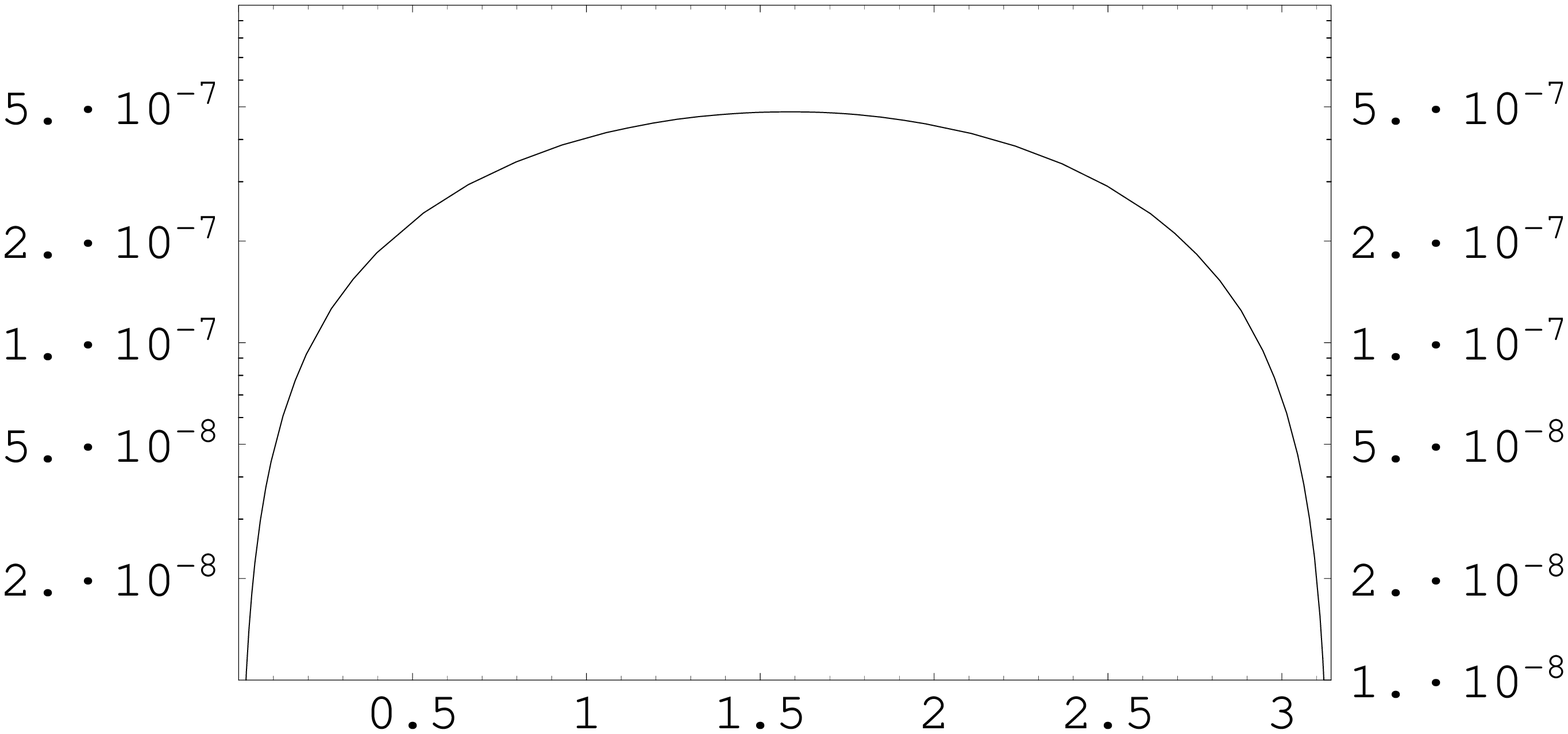}}
\put(5.5,0.3){\footnotesize $\theta$}
\put(-0.8,3.5){\footnotesize $|J|$}
\end{picture}
\end{center}
\caption{The Jarlskog determinant in the case of $R = 3$, $\alpha=0.5$, $|r_{H}| = 0.2$, $|r_{H'}| = 0.45$, $r_{Q} = 0.2$, $r_{U} = 0.2$ and  $r_{D} = 0.2$}
\label{fig5}
\end{figure}

To see the magnitude of the $CP$-violating phase, we calculate the Jarlskog determinant $J$\cite{rf:JA} which is the rephaseing-invariant parameter.
In Fig. 5, we give a result for $J$ in the parameter region which gives 
above CKM matrix.
Here, $\theta$  defined in  (\ref{eq:phase-r}) is $\theta = - \arg [r_{Q}] - 3 \arg [ r_{H} ] $.
The magnitude of $J$, $|J| \sim {\cal O}( 10^{-7})$, looks much smaller than 
what is expected in the SM, $ |J| \sim {\cal O }( 10^{-5})$.

Finally,  let us mention on masses of charged leptons.
Since charged leptons acquire masses through Yukawa couplings  to Higgs $h'$,
their mass ratios are given by those of down-type quarks with the replacement
$r_{Q} \rightarrow r_{L}$ and $r_{D} \rightarrow r_{E}$.
As show in Fig. 2, mass ratios are almost insensitive to these parameters.
Therefore the present minimal model predicts essentially equivalent mass ratios
for charged leptons and down-type quarks.
This prediction is unfavorable to the observation.
The model cannot reproduce both of the mass ratios simultaneously.

\section{Summary and Discussions}

In this paper we have proposed a new mechanism  for ``spontaneous generation of generations'' based on the $SU(1,1)$ gauge symmetry. 
This mechanism works well in the grand unified theories.
We can naturally understand the coexistence of chiral massless particles and
super heavy GUT partners without fine tuning.
At the same time, this mechanism yields the hierarchal Yukawa interactions 
among chiral generations naturally.

We investigated the characteristic feature of the model through the numerical analysis under the minimal parameter set which is motivated by the grand unification and also by the economical purpose.
We found that the observed hierarchies of both up-type and down-type quarks are qualitatively reproduced under the reasonable values of the parameters.
This minimal model is also capable to realize the characteristic structure of 
the CKM matrix.

The most serious difficulty of the minimal model is that the model cannot simultaneously reproduced the mass ratios of down-type quarks and charged leptons.
The smallness of the predicted value of $J$ may be also the problem if we seek the origin of the $CP$-violation at the phase of the CKM matrix.

From this survey, we conclude that the minimal model based on the severely restricted set of parameters does not have enough capacity to reproduce the reality.
The model must be modified by some relaxation of the constraints.
At this stage, the compatibility of the model with the underling grand unification symmetry will give useful guiding principle.
We expect that the further attempts along this line will give us deeper insight on the generations of quarks and leptons.

\section*{Acknowledgements}

 The work of K.~I. was supported in part by the Grant-in-Aid of the Ministry of
Education, Science, Sports and Culture, Government of Japan (No. 11640282),
and Priority area ``Supersymmetry and Unified Theory of Elementary Particles''
 (No. 707).


\begin{thebibliography}{99}


\bibitem{rf:horizon}
T.~Maehara and T.~Yanagida, \PTP{61,1979,1434}.\\
F.~Wilczek and A.~Zee, \PL{42,1979,421}.\\
C.~D.~Froggatt and H.~B.~Nielsen, \NP{B147,1979,277}.
\bibitem{rf:KI1}
K.~Inoue, \PTP{93,1995,403}.
\bibitem{rf:KI2}
K.~Inoue, \JL{Prog.~Theor.~Phys.~Suppl.,123,1996,319}.
\bibitem{rf:RL}
F.~A.~Wilczek,~A.~Zee,~R.~L.~Kingsley and S.~B.~Treiman, \PR{D12,1975,2768}.\\
A.~De.~R\'{u}jula,~H.~Georgi and S.~L.~Glashow, \PR{D12,1975,3589}.\\
H.~Fritzsch,~M.~Gell-Mann and P.~Minkowski, \PL{59B,1975,256}.\\
K.~Inoue,~A.~Kakuto and Y.~Nakano, \PTP{58,1977,630}.\\ 
M.~Yoshimura, \PTP{58,1977,972}.
\bibitem{rf:KAZUO}
K.~Fujikawa,
hep-ph/9604358.\\
K.~Fujikawa, \PTP{92,1994, 1149}.
\bibitem{rf:MSSM}
P.~Fayet, \PL{69B,1977,489}\\
K.~Inoue,~A.~Kakuto,~H.~Komatu and S.~Takesita, \PTP{67,1982,1889}; \PTP{68,1982,927}.\\
L.~E.~Ib\'{a}\~{n}ez and G.~G.~Ross, \PL{110B,1982,215}.\\
L.~Alvarez-Gaume,~J.~Polchinski and M.~K.~Wise, \NP{B221,1983,495}.\\
J.~Ellis,~J.~S.~Hagelin,~D.~V.~Nanopoulos and K.~Tamvakis, \PL{125B,1983,275}.
\bibitem{rf:GUT}
H.~Georgi and S.~L.~Glashow, \PRL{32,438,1974}.
\bibitem{rf:SUSYGUT}
S.~Dimopoulos and H.~Georgi, \NP{B193,1981,150}.\\
N.~Sakai, \JL{Z.~Phys.,C11,1982,153}.
\bibitem{rf:CKM}
N.~Cabibbo, \PRL{10,531,1963}.\\
M.~Kobayashi and K. Maskawa, \PTP{49,282,1972}.
\bibitem{rf:JA}
C.~Jarlskog, \PRL{55,1984,1039}.
\end{thebibliography}
\end{document}